\documentclass[journal,10pt]{IEEEtran}
\usepackage{graphicx}
\usepackage{epstopdf}
\usepackage{multirow}
\usepackage{amsmath}
\usepackage{mathtools}
\usepackage{authblk}
\usepackage{cite}
\usepackage{cleveref}
\usepackage{latexsym}
\usepackage[justification=centering]{caption}
\usepackage{amsthm}
\usepackage{balance}
\usepackage{relsize}
\usepackage[nodisplayskipstretch]{setspace}
\usepackage{color}
\usepackage{amssymb}
\usepackage{eucal}
\usepackage{url}
\usepackage{booktabs,subcaption,amsfonts,dcolumn}
\newtheorem{prop}{Proposition}
\usepackage[nodisplayskipstretch]{setspace}
\usepackage{color}
\newcolumntype{d}[1]{D..{#1}}
\usepackage{lipsum}

\usepackage[nodisplayskipstretch]{setspace}
\usepackage{color}
\definecolor{Mypink}{RGB}{255,0,255}
\definecolor{Myorange}{RGB}{255,102,0}
\definecolor{Mygreen}{RGB}{0,153,0}
\definecolor{Myblue}{RGB}{0,0,255}

\DeclareMathAlphabet\mathbfcal{OMS}{cmsy}{b}{n}
\begin{document}

\title{Non-Orthogonal Multiple Access in the Presence of Additive Generalized Gaussian Noise}

\renewcommand\Authfont{\fontsize{12}{14.4}\selectfont}
\renewcommand\Affilfont{\fontsize{9}{10.8}\itshape}

\author{
Lina~Bariah,~\IEEEmembership{Member,~IEEE,}
        Sami~Muhaidat,~\IEEEmembership{Senior~Member,~IEEE,}
        Paschalis~C. Sofotasios,~\IEEEmembership{Senior~Member,~IEEE,}
				Sanjeev~Gurugopinath,~\IEEEmembership{Member,~IEEE,}
				Walaa~Hamouda,~\IEEEmembership{Senior~Member,~IEEE,}
				and Halim~Yanikomeroglu,~\IEEEmembership{Fellow,~IEEE,}
				\thanks{L. Bariah and S. Muhaidat are with the KU Center for Cyber-Physical Systems, Department of Electrical and Computer Engineering, Khalifa University, Abu Dhabi 127788, UAE, (e-mails: {lina.bariah, muhaidat}@ieee.org).}
				\thanks{P. C. Sofotasios is with the Center for Cyber-Physical Systems, Department of Electrical and Computer Engineering, Khalifa University, Abu Dhabi 127788, UAE, and also with the Department of Electrical Engineering, Tampere University, Tampere 33101, Finland (e-mail: p.sofotasios@ieee.org).}
				\thanks{S. Gurugopinath is with the Department of Electronics and Communication Engineering, PES University, Bengaluru 560085, India (e-mail: sanjeevg@pes.edu).}
				\thanks{W.  Hamouda  is  with  the  Department  of  Electrical  and  Computer  Engineering,  Concordia  University,  Montreal,  QC,  H3G  1M8,  Canada  (e-mail:hamouda@ece.concordia.ca).}
				\thanks{H. Yanikomeroglu is with the Department of Systems and Computer Engineering, Carleton University, Ottawa, ON K1S 5B6, Canada (e-mail: halim@sce.carleton.ca).}
		\vspace{-0.7cm}		}

\maketitle

\begin{abstract}
\boldmath
In this letter, we investigate the performance of non-orthogonal multiple access (NOMA), under the assumption of generalized Gaussian noise (GGN), over Rayleigh fading channels. Specifically, we consider a NOMA system with $L$ users, each of which is equipped with a single antenna, and derive an exact expression for the pairwise error probability (PEP). The derived PEP expression is subsequently utilized to derive a union bound on the bit error rate (BER) and to quantify the diversity orders realized by NOMA users in the presence of additive white (AW) GGN. Capitalizing on the derived PEP expression and the union bound, the error rate performance of NOMA users is further evaluated for different special cases of AWGGN. The derived analytical results, corroborated by simulation results, show that the shaping parameter of the GGN $(\alpha)$ has negligible effect on the diversity gains of NOMA users, particularly for large $\alpha$ values. Accordingly, as in the case of additive white Gaussian noise (AWGN), the maximum achievable diversity order is determined by the user's order. 
\end{abstract} 

\begin{keywords}
Additive generalized Gaussian noise, diversity order, NOMA, pairwise error probability.
\end{keywords}

\vspace{-0.2cm}

\IEEEpeerreviewmaketitle
\section{Introduction}

The rapid increase in the number of connected devices and the explosive growth of mobile traffic, which is expected to reach 77.5 Exabytes per month in 2022, have imposed stringent requirements on the 5th generation (5G) of wireless networks, such as massive connectivity, low latency and enhanced spectral efficiency \cite{Cisco}. Consequently, in order to address these challenges, new communication paradigms have been recently proposed in the literature, including the proposal of new multiple access techniques. More specifically, non-orthogonal multiple access (NOMA)  has recently emerged as a key enabling technology for 5G wireless networks \cite{6868214}. NOMA was included in the third generation partnership project long-term evolution (3GPP LTE) Advanced Release 13 in order to realize multiuser superposition transmission (MUST) \cite{3gpp}. 


Recently, several variants of NOMA have been proposed, which can be generally classified into two categories, namely, power domain multiplexing \cite{6692652} and code domain multiplexing \cite{4471881}. In code domain multiplexing, e.g., multiple access with low-density spreading (LDS), user-specific spreading codes are utilized in order to multiplex different users in the code domain. On the other hand, in power domain multiplexing, signals of different users are multiplexed in the power domain by assigning different power levels to different users. At the users’ terminals, multi-user detection is realized by successive interference cancellation (SIC). 

The superiority of NOMA over conventional orthogonal multiple access techniques has been demonstrated  in several aspects in the recent literature. This includes spectral efficiency, in which multiple users are served using the same time and frequency resources, interference mitigation through SIC, and the support for massive connectivity. 

The current literature on NOMA  has primarily focused on the additive white Gaussian noise (AWGN) assumption, which essentially represents the thermal noise at the receiver side \cite{8636496,8705669,7870569,7398134}. Nevertheless, in many practical scenarios, the AWGN assumption is not sufficient to accurately describe the noise model as it ignores other sources of noise. Specifically, it has been shown that the combined statistics of interference and noise in ultra wide-band communication systems follows the generalized Gaussian distribution (GGD) \cite{6185516}. 

In underwater acoustic (UWA) systems, the noise sources are limited in number. Thus, the noise of UWA systems is modeled as impulsive noise. Furthermore, the noise model in power line communications (PLC) is characterized by a mixture of Gaussian and Laplacian noise \cite{r1}, requiring an accurate and unified model to represent such distinct noise model. 
It is worth noting that the impulsive noise component in the simplified class A noise model, which is modeled by Laplacian distribution \cite{8404425}, and the Gaussian distribution are special cases of the GGD.

We emphasize that the impact of additive white generalized Gaussian noise (AWGGN) on the performance of NOMA is not comprehensively understood yet, since it has not been addressed in the related open literature, which calls for a thorough investigation. We note that such an investigation is indeed compelling for the successful realization of NOMA in some particular applications and for determining the actual performance limits, particularly  in terms of the system's reliability.

Motivated by the above, in this letter, we analyze the performance of a NOMA system over Rayleigh fading channels in the presence of AWGGN. In particular, we derive a novel exact closed-form expression for the pairwise error probability (PEP), which is subsequently used to quantify the diversity order of NOMA users under the considered scenario. To the best of the authors' knowledge, such a performance study has not been reported in the open literature. 

\section{System Model} 
\label{sec:model}

We consider a downlink NOMA system which consists of a single base station (BS) and $L$ ordered users, $U_{1}, \cdots, U_{L}$, with asymmetric channel gains. Hence, users are ordered based on their channel gains, i.e., $h_{1}<h_{2}<\cdots<h_{L}$, where $h_{l}$ is the fading envelope between the BS and the $l$th user, which follows the Rayleigh distribution with zero mean and unit variance, i.e., $\sigma_{l}^{2} = 1$. We further assume that all channel coefficients, $h_{i}$, $i \in \{ 1,\cdots,L\}$, are independent and identically distributed. Following the key principle of NOMA, users are allocated different power levels depending on their channel gains, i.e., higher power coefficients are assigned to far users, while near users are allocated lower power coefficients. In power domain NOMA, the transmitted signals of the $L$ users are multiplexed in the power domain, yielding 	\vspace{-0.2cm}
\begin{equation}\vspace{-0.2cm}
s = \sum_{i=1}^{L}\sqrt{a_{i}P}x_{i} 
\end{equation}  
where $a_{i}$ and $x_{i}$ are the power allocation coefficient and the transmitted symbol of the $i$th user, respectively, and $P$ is the total transmission power at the BS. Accordingly, following the commonly used mathematical signal model \cite{6185516,porakis}, the received signal at the $l$th user is given by 
\begin{equation}\vspace{-0.2cm}
\label{eq:y}
y_{l} = h_{l}s+n_{l} 
\end{equation}
where the noise term, $n_{l} \in \mathbb{R}$, represents the AWGGN with zero mean and variance $N_{0}/2$. Specifically, the probability density function (PDF) of $n_{l}$ is given by \cite{6620186} 
\begin{equation}
\label{eq:pdf}
f(n_{l})=\frac{\alpha \Lambda}{2 \Gamma(1/\alpha)}\, \textup{exp} \left ( -\Lambda^{\alpha}\left | n_{l} \right |^{\alpha} \right ) 
\end{equation} 
where $\Gamma(.)$ is the complete gamma function, $\alpha \in \mathbb{R}^{+}$ denotes the shaping parameter and $\Lambda = 2\Lambda_{0}/N_{0}$ is the noise power normalization coefficient, $\Lambda_{0} = \Gamma(3/\alpha)/\Gamma(1/\alpha)$. It is worth noting that most of the common noise models are considered as special cases of the AWGGN. For example, when $\alpha = 2$, the PDF in \eqref{eq:pdf} reduces to the Gaussian noise PDF \cite{porakis}, while the Laplacian noise model can be obtained by setting $\alpha$ to 1 \cite{8404425}. Note that from classical information-theoretic perspective, for a fixed noise variance, Gaussian noise has been shown to be the worst-case additive noise for wireless channels. This follows from the fact that the Gaussian distribution maximizes the entropy, and thus, can be considered as a lower bound on the channel capacity \cite{6477131}. Hence, the practical values of $\alpha$ is assumed to be less than 2, i.e., $\alpha \leq 2$.  

At the users’ terminals, SIC is utilized to realize multi-user detection and mitigate interference \cite{8732367}. Particularly, user $U_l$, $l = 2,\cdots, L$, first detects users' signals with higher power coefficients, $U_j$  $(1  \leq j \leq l-1)$, and then subtracts them from its received signal. Next, user $l$ detects its own signal by treating users with lower power coefficients, $U_k$ $(k > l)$, as noise. Consequently, after performing SIC, the received signal at $U_l$ can be written as 
\begin{equation}
\label{eq:yprime}
y'_{l} = h_{l}\left ( \sqrt{a_{l}\bar{\gamma}}\, x_{l} + X\right )+\tilde{n}_{l} 
\end{equation} 
where $\tilde{n}_{l}$ is the normalized AWGGN with zero mean and unit variance, $\bar{\gamma} = 2P/N_{0}$ is the average transmit signal-to-noise ratio (SNR), and 
\begin{equation}
\label{eq:X}
X = \underbrace{\sum_{i=1}^{l-1}\sqrt{a_{i}\bar{\gamma}}\, \hat{\Delta}_{i}}_{\textup{SIC}} +\underbrace{\sum_{j=l+1}^{L}\sqrt{a_{j}\bar{\gamma}}\, x_{j}}_{\textup{IUI}}.   
\end{equation}
\noindent Here, $\hat{\Delta}_{i}=x_i - \hat{x}_i$ denotes the error signal at the $i$th layer of SIC, given that $\hat{x}_i$ is the detected symbol of the $i$th user. The residual interference from higher order users is treated as additive noise, which has less impact on the error rate performance due to their reduced power levels. It is worth mentioning that, as the value of the shaping parameter ($\alpha$) increases, the GGD tail becomes tighter, which means the noise level becomes lower. Subsequently, as $\alpha$ value decreases, the effect of AWGGN on the received signal becomes higher. After the SIC process, user $l$ performs maximum likelihood decoding to detect its own signal. Based on the noise model in \eqref{eq:pdf} and assuming the effect of multi-user interference is negligible, the received signal $y'_{l}$, conditioned on $h_{l}$ and $x_{l}$, is modeled as GG with mean $=h_{l}\left ( \sqrt{a_{l}\bar{\gamma}}\, x_{l}\right )$ and unit variance. Therefore, the conditional PDF of $y'_l$ can be expressed as 
\begin{equation}
f(y'_{l}|x_{l},h_{l}) = \frac{\alpha \Lambda}{2 \Gamma(1/\alpha)}\, \textup{e}^{ -\Lambda^{\alpha}\left | y'_{l} - h_{l}\left ( \sqrt{a_{l}\bar{\gamma}}\, x_{l}\right )\right |^{\alpha} }. \vspace{-0.2cm}
\end{equation}
The receiver detects the transmitted signal according to the following ML criterion \cite{6185516} \vspace{-0.2cm}
\begin{equation}
\hat{x}_{l} = \textup{arg}\; \underset{\tilde{x} \in \phi}{\textup{max}} f(y'_{l}|x_{l},h_{l}) \vspace{-0.2cm}
\end{equation}\vspace{-0.2cm}
which can be further simplified to
\begin{equation}
\hat{x}_{l} = \textup{arg}\; \underset{\tilde{x} \in \phi}{\textup{min}}\left | {y}'_{l}-\sqrt{a _{l}\bar{\gamma}}\;h_{l}\tilde{x} \right | 
\end{equation} 
where $\phi$ is an arbitrary signal constellation set.
\vspace{-0.2cm}
\section{Performance Analysis} 
\label{sec:pep}
\subsection{Pairwise Error Probability Analysis} 
\vspace{-0.1cm}
In this section, we derive a novel closed-form expression for the PEP for all NOMA users subject to AWGGN. The PEP is defined as the probability of erroneously detecting symbol $\check{x}_{l}$ when symbol $x_{l}$ is transmitted. It is worth mentioning that the PEP constitutes the basic building block for the derivation of the union bound on the error probability.
\begin{prop}
\label{theorem:pep}
The exact PEP expression of the $l$th user in the NOMA scheme can be expressed by \eqref{eq:pep1}, on the top of the next page, where $\mu = 1$ for $\upsilon < 0$ and $\mu = 0$ for $\upsilon > 0$. In \eqref{eq:pep1}, $\check{\Delta}_{l}=x_{l}-\check{x}_{l}$ denotes the error signal of the $l$th user. Also,  
\begin{figure*}[ht]
\begin{equation}
\label{eq:pep1}
\begin{split}
\textup{Pr}\left ( x_{l},\check{x}_{l} \right )=\frac{A_{l}}{2 \Gamma(\frac{1}{\alpha})}\sum_{i=0}^{l-1}\binom{l-1}{i}&\frac{\left ( -1 \right )^{i}}{\delta_{l,i}}\Bigg [ \Gamma\left ( \frac{1}{\alpha} \right )+\frac{(-1)^\mu \sqrt{k}\alpha \sqrt{\Lambda_{0}}\left | \upsilon \right |\pi}{\sqrt{a_{l}\bar{\gamma}\delta_{l,i}} \Re \left \{ \check{\Delta}_{l}\right \}(2 \pi)^{\frac{k}{2}(\frac{\alpha}{2}+1)}} \\
&  \times G^{k,\frac{k \alpha}{2}}_{\frac{k \alpha}{2},k}\left ( \frac{1}{k^k}\left ( \frac{\sqrt{\Lambda_{0}} \left | \upsilon \right |}{\Re \left \{ \check{\Delta}_{l}\right \}} \right )^{\alpha k} \left ( \frac{\alpha k}{2 a_{l} \bar{\gamma}\delta_{l,i}} \right )^{\frac{\alpha k}{2}}\middle\vert  \begin{matrix}
\frac{2(1-\frac{1}{2})}{\alpha k},\cdots, \frac{2(\frac{\alpha k}{2}-\frac{1}{2})}{\alpha k} \\ 
0,\cdots,\frac{k-1}{k}\end{matrix} \right ) \Bigg ]. \;\; \begin{matrix}
\end{matrix} 
\end{split}
\end{equation}
\hrulefill
\vspace*{4pt}
\end{figure*}
\begin{equation}
\delta_{l,i}=\left ( L-l+1+i \right )
\end{equation}
\begin{equation}
A_{l} = L!/[(l-1)!(L-l)!]
\end{equation}
\begin{equation} 
\upsilon = [\textup{mod}(X)]^2-[\textup{mod}(\zeta)]^{2}
\end{equation}\vspace{-0.2cm}
\begin{equation}
\label{eq:zeta}
\zeta = \sqrt{a_{l}\bar{\gamma}}\check{\Delta}_{l}+X
\end{equation}
\noindent where $\textup{mod}(b)$ denotes the modulus of a complex number $b$ and the value of $k$ is selected such that $t = k \alpha / 2$ reduces to an integer number. 
\end{prop} 
\vspace{-0.4cm}
\begin{proof}
Following the basic definition of the PEP, and given that $x_{l}$ and $\check{x}_{l}$ are the respective transmitted and incorrectly detected symbols of the $l$th user, the conditional PEP can be written as follows 
\begin{equation}
\label{eq:pep2}
\textup{Pr}\left ( x_{l},\check{x}_{l}\mid h_{l} \right )= \textup{Pr}\left ( \left | h_{l}\zeta+\tilde{n}_{l} \right |^{2} \leq \left | h_{l}X+\tilde{n}_{l} \right |^{2} \right ) 
\end{equation}
where the inter-user interference and the effect of imperfect SIC are represented by $X$ and $\zeta$, which are defined in \eqref{eq:X} and \eqref{eq:zeta}, respectively. After some mathematical manipulations, the conditional PEP in \eqref{eq:pep2} can be written as 
\begin{equation}
\label{eq:pep3}
\textup{Pr}\left ( x_{l},\check{x}_{l}\mid h_{l} \right )= \textup{Pr}\left ( \underbrace{2\Re \left \{ \sqrt{a_{l}\bar{\gamma}}h_{l}\check{\Delta}_{l}\tilde{n}_{l} \right \} }_{N} \leq \left | h_{l} \right | ^{2} \upsilon \right ) 
\end{equation}
\noindent where $\Re \left \{ z \right \}$ denotes the real part of the complex variable $z$. Note that for the special case of binary phase shift keying (BPSK) modulation, $N$ in \eqref{eq:pep3} reduces to $N=2\sqrt{a_{l}\bar{\gamma}}h_{l}\check{\Delta}_{l}\tilde{n}_{l}$. Since $\tilde{n}_{l}$ represents the normalized AWGGN with zero mean and unit variance, the decision variable $N$ follows the GGD with zero mean and variance, $\sigma_{N}^{2}= 2a_{l}\bar{\gamma} h_{l}^{2}\left [\Re \left \{ \check{\Delta}_{l}\right \} \right ]^{2}$. Therefore, the PDF of $N$ can be obtained from \eqref{eq:pdf}, after substituting $\Lambda$ by  
\begin{equation} 
\label{eq:lambda}
\lambda= \sqrt{\Lambda_{0}}/[\sqrt{2a_{l}\bar{\gamma}} h_{l} \Re \left \{ \check{\Delta}_{l}\right \}]. 
\end{equation}  
Consequently, the conditional PEP can be evaluated as  
\begin{equation}
\label{eq:pep4}
\begin{split}
\textup{Pr}\left ( x_{l},\check{x}_{l}\mid h_{l} \right )&= \textup{Pr}\left ( N \leq  h_{l}^{2}\upsilon \right ) \\
&= \int_{-\infty}^{ h_{l}^{2}\upsilon}\frac{\alpha \lambda}{2 \Gamma(1/\alpha)}\, \textup{exp} \left ( -\lambda^{\alpha}\left | z \right |^{\alpha} \right )dz. 
\end{split}
\end{equation} 
By noting that the value of $\upsilon$ is determined based on the error codewords, $\upsilon$ has two cases, namely, $\upsilon>0$, which denotes destructive interference, and $\upsilon<0$ denoting constructive inference. Consequently, the integral in \eqref{eq:pep4} can be re-written as \eqref{eq:pep5}, on the top of the next page.\\
\begin{figure*}[ht]
\begin{equation}
\label{eq:pep5}
\textup{Pr}\left ( x_{l},\check{x}_{l}\mid  h_{l}  \right )=  \begin{cases}
\frac{\alpha \lambda}{2 \Gamma(1/\alpha)}\left [\int_{0}^{\infty} \textup{exp} \left ( -\lambda^{\alpha} z^{\alpha} \right )dz+\int_{0}^{ h_{l}^{2}\left |\upsilon  \right |} \textup{exp} \left ( -\lambda^{\alpha} z^{\alpha} \right )dz  \right ], & \upsilon>0\\ \\
\frac{\alpha \lambda}{2 \Gamma(1/\alpha)}\left [\int_{ h_{l}^{2}\left |\upsilon  \right |}^{\infty} \textup{exp} \left ( -\lambda^{\alpha} z^{\alpha} \right )dz  \right ], & \upsilon<0 
\end{cases}
\end{equation} 
\hrulefill
\vspace*{4pt}
\end{figure*}
The three integrals in \eqref{eq:pep5} can be evaluated using the following \cite[Eqs. 3.381.8, 3.381.9, 3.381.10]{IntTable} 
\begin{equation}
\int_{0}^{u}r^{m}\textup{e}^{-\beta r^{n}}dr=\frac{\gamma\left ( \varepsilon  ,\beta u^{n}\right )}{n \beta^{\varepsilon}}
\end{equation}
\begin{equation}
\int_{u}^{\infty}r^{m}\textup{e}^{-\beta r^{n}}dr=\frac{\Gamma\left ( \varepsilon  ,\beta u^{n}\right )}{n \beta^{\varepsilon}}
\end{equation}
and
\begin{equation}
\int_{0}^{\infty}r^{m}\textup{e}^{-\beta r^{n}}dr=\frac{\gamma\left ( \varepsilon  ,\beta u^{n}\right )+\Gamma\left ( \varepsilon  ,\beta u^{n}\right )}{n \beta^{\varepsilon}}
\end{equation} 
yielding \eqref{eq:pep6}, at the top of the next page, where $\varepsilon = (m+1)/n$. 
\begin{figure*}[ht]
\begin{equation}
\label{eq:pep6}
\textup{Pr}\left ( x_{l},\check{x}_{l}\mid h_{l}   \right )=  
\frac{1}{2 \Gamma(1/\alpha)}\left [\Gamma\left ( \frac{1}{\alpha} \right ) + (-1)^\mu \gamma\left ( \frac{1}{\alpha},\lambda^{\alpha} h_{l}^{2\alpha}\left | \upsilon \right |^{\alpha} \right )  \right ]. 
\end{equation} 
\hrulefill
\vspace*{4pt}
\end{figure*}
In \eqref{eq:pep6}, $\gamma\left ( A,z\right )$ denotes the lower incomplete gamma function. Finally, to obtain the unconditional PEP expression, we integrate the conditional PEP in \eqref{eq:pep6} over the PDF of $\omega_{l} \triangleq \left |h_{l}  \right |$. Recalling that users are ordered based on their channel gains and the channel gain variance of the $l$th user is normalized to unity, the PDF of $\omega_{l}$ can be expressed as follows \cite{8501953} \vspace{-0.3cm}
\begin{equation} \vspace{-0.2cm}
\label{eq:pdf3}
f_{l}\left ( \omega_{l} \right ) = A_{l}\omega_{l}\sum_{i=0}^{l-1}\binom{l-1}{i}\left ( -1 \right )^{i} \textup{exp}\left ( - \frac{\delta_{l,i}\omega_{l}^{2}}{2} \right ). 
\end{equation}
Therefore, the unconditional PEP can be evaluated as 
\begin{equation}
\label{eq:pep7}
\textup{Pr}\left ( x_{l},\check{x}_{l} \right )=\frac{A_{l}}{2 \Gamma(1/\alpha)}\sum_{i=0}^{l-1}\binom{l-1}{i}\left ( -1 \right )^{i}\left [T_{1}+(-1)^\mu T_{2}  \right ],
\end{equation}
\vspace{-0.4cm}where 
\begin{equation}
\label{eq:T1}
T_{1} = \Gamma\left ( \frac{1}{\alpha} \right )\int_{0}^{\infty }\omega_{l}\,\textup{exp}\left ( -\frac{\delta_{l,i}\omega_{l}^{2}}{2} \right )d\omega_{l} 
\end{equation}
and \vspace{-0.2cm}
\begin{equation}  \vspace{-0.2cm}
\label{eq:T2}
\begin{split}
T_{2} = \int_{0}^{\infty }\omega_{l}\,\textup{exp}\left ( -\tfrac{\delta_{l,i}\omega_{l}^{2}}{2} \right ) \gamma\left ( \tfrac{1}{\alpha},\tfrac{[\sqrt{\Lambda_{0}}]^{\alpha}\omega_{l}^{\alpha}\left | \upsilon \right |^{\alpha}}{\left [2a_{l}\bar{\gamma}  \right ]^{\frac{\alpha}{2}}\left [\Re \left \{ \check{\Delta}_{l}\right \} \right ]^{\alpha}} \right )d\omega_{l}.
\end{split}
\end{equation}
The integral in \eqref{eq:T1} can be obtained in closed-form as \vspace{-0.2cm}
\begin{equation} \vspace{-0.2cm}
\label{eq:T1a}
T_{1} = \Gamma\left ( 1/\alpha \right )/\delta_{l,i}.
\end{equation}
Utilizing integration by parts, the integral in \eqref{eq:T2} can be simplified to
\begin{equation}\vspace{-0.2cm}
\label{eq:T2b}
\begin{split}
T_{2} = &\frac{\alpha \sqrt{\Lambda_{0}} \left | \upsilon \right |}{\sqrt{2a_{l}\bar{\gamma}}\Re \left \{ \check{\Delta}_{l}\right \}\delta_{l,i}}\int_{0}^{\infty }\textup{exp}\left ( -\frac{[\sqrt{\Lambda_{0}}]^{\alpha}\omega_{l}^{\alpha}\left | \upsilon \right |^{\alpha}}{\left [2a_{l}\bar{\gamma}  \right ]^{\frac{\alpha}{2}}\left [\Re \left \{ \check{\Delta}_{l}\right \} \right ]^{\alpha}} \right ) \\
& \;\;\;\;\;\;\;\;\;\;\;\;\;\;\;\;\;\;\;\;\;\;\;\;\;\;\;\;\; \times \textup{exp}\left ( -\frac{\delta_{l,i}\omega_{l}^{2}}{2} \right ) d\omega_{l}.
\end{split}
\end{equation} 
Using the Meijer's G-function $G^{m,n}_{p,q}\left ( .\mid. \right )$ representation of the exponential function \cite{8539989} and by setting $\theta=\omega_{l}^{2}$, the integral in \eqref{eq:T2b} can be rewritten as \vspace{-0.2cm} 
\begin{equation}\vspace{-0.2cm}
\label{eq:T2c}
\begin{split}
&T_{2} = \frac{\alpha \sqrt{\Lambda_{0}} \left | \upsilon \right |}{2\sqrt{2a_{l}\bar{\gamma}} \Re \left \{ \check{\Delta}_{l}\right \}\delta_{l,i}}\int_{0}^{\infty }\frac{1}{\sqrt{\theta}}\\
& \times G^{1,0}_{0,1}\left ( \tfrac{\delta_{l,i}\theta}{2}\Bigm| \begin{matrix}
-\\ 0
\end{matrix} \right ) G^{1,0}_{0,1}\left ( \frac{[\sqrt{\Lambda_{0}}]^{\alpha}\theta^{\frac{\alpha}{2}}\left | \upsilon \right |^{\alpha}}{\left [2a_{l}\bar{\gamma}  \right ]^{\frac{\alpha}{2}} \left [\Re \left \{ \check{\Delta}_{l}\right \} \right ]^{\alpha}}\Bigm| \begin{matrix}
-\\ 0
\end{matrix} \right )d\theta
\end{split}
\end{equation}
\vspace{-0.2cm}which can be evaluated, using \cite[Eq. 2.24.1.1]{int22}, as 
\begin{equation}
\label{eq:T2F}
\begin{split}
&T_{2} = \tfrac{\sqrt{k}\alpha \sqrt{\Lambda_{0}} \left | \upsilon \right |\pi}{\sqrt{a_{l}\bar{\gamma}\delta_{l,i}}\Re \left \{ \check{\Delta}_{l}\right \}(2 \pi)^{\tfrac{k}{2}(\tfrac{\alpha}{2}+1)}} \times \\
& G^{k,\tfrac{k \alpha}{2}}_{\tfrac{k \alpha}{2},k}\left ( \tfrac{1}{k^k}\left ( \tfrac{\sqrt{\Lambda_{0}} \left | \upsilon \right |}{\Re \left \{ \check{\Delta}_{l}\right \}}\sqrt{\tfrac{\alpha k}{2 a_{l} \bar{\gamma}\delta_{l,i}}} \right )^{\alpha k} \middle\vert  \begin{matrix}
\tfrac{2(1-\tfrac{1}{2})}{\alpha k},.., \frac{2(\tfrac{\alpha k}{2}-\tfrac{1}{2})}{\alpha k} \\ 
0,..,\tfrac{k-1}{k}\end{matrix} \right ). 
\end{split}
\end{equation}
Finally, substituting \eqref{eq:T1a} and \eqref{eq:T2F} in \eqref{eq:pep7} yields the unconditional PEP expression given by \eqref{eq:pep1}.
\end{proof} 
In the following, to gain some insights into the system performance under AWGGN, we consider two  special cases, namely,  $\alpha = 1$ and $\alpha = 2$, and derive the corresponding PEP expressions. 
\\
\textbf{Case 1 ($\alpha = 1$):} In this case, we introduce a simplified PEP expression for the special case of $\alpha = 1$, which represents the Laplacian noise PDF. With $\alpha =1$, the Meijer's G-function in \eqref{eq:pep1} can rewritten as 
\begin{equation}
\label{eq:MG}
\begin{split}
G^{2,1}_{1,2}\left (   \tau^{2} \middle\vert  \begin{matrix}
\frac{1}{2} \\ 
0,\frac{1}{2}\end{matrix} \right )= \pi \textup{exp}\left ( \tau^{2} \right )\textup{erfc}\left ( \tau \right )
\end{split}
\end{equation}  
where 
\begin{equation}
\tau = \frac{\sqrt{\Lambda_{0}} \left | \upsilon \right |}{2\sqrt{a_{l}\bar{\gamma}\delta_{l,i}} \Re \left \{ \check{\Delta}_{l}\right \}}.
\end{equation}
Hence, the PEP of the $l$th user, under this scenario, can be obtained as \vspace{-0.2cm}
\begin{equation}
\label{eq:PEP_L}
\begin{split}
\textup{Pr}\left ( x_{l},\check{x}_{l} \right )=\tfrac{A_{l}}{2 }\sum_{i=0}^{l-1}\binom{l-1}{i}\tfrac{\left ( -1 \right )^{i}}{\delta_{l,i}} \Bigg [ 1+(-1)^\mu \tau & \sqrt{\pi}  \textup{exp}\left ( \tau^2 \right ) \\ & \times \textup{erfc}\left ( \tau \right ) \Bigg ].
\end{split}
\end{equation} 
\vspace{-0.5cm}
\\
\textbf{Case 2 ($\alpha = 2$):} In this scenario, the AWGGN simplifies to AWGN.  Hence, by setting $k$ to 1, the Meijer's G-function in \eqref{eq:pep1}  can be written  as 
\begin{equation}
\label{eq:MG1}
\begin{split}
G^{1,1}_{1,1}\left (  4\tau^{2}  \middle\vert  \begin{matrix}
0.5 \\ 
0\end{matrix} \right )= \Gamma\left ( 0.5 \right )\left ( 4\tau^{2}+1 \right )^{-1/2}.
\end{split}
\end{equation}  
Therefore,  the PEP of the $l\rm{th}$ user can be expressed as \vspace{-0.2cm}
\begin{equation}
\label{eq:PEP_L1}
\begin{split}
\textup{Pr}\left ( x_{l},\check{x}_{l} \right )=\frac{A_{l}}{2 }\sum_{i=0}^{l-1}\binom{l-1}{i}\frac{\left ( -1 \right )^{i}}{\delta_{l,i}}\Bigg [ 1 + (-1)^\mu \frac{2\tau}{ \sqrt{4\tau^{2}+1}} \Bigg ]. 
\end{split}
\end{equation} \vspace{-0.2cm}
\vspace{-0.8cm}
\subsection{Union Bound on the BER Performance}
It is widely accepted that PEP provides an indispensable tool for the derivation of union bounds on the bit error rate (BER) performance of digital communication systems. Recalling that $x_{l}$ and $\check{x}_{l}$ denote the transmitted and the incorrectly decoded symbols of the $l$th user, the BER union bound can be written as \cite{4524293}
\vspace{-0.2cm} 
\begin{equation}\vspace{-0.2cm} 
\label{eq:UB}
P_{UB} \leq \frac{1}{q} \sum_{x_{l}}\textup{Pr}\left ( x_{l} \right )\sum_{x_{l} \neq \check{x}_{l}}e\left ( x_{l}\rightarrow  \check{x}_{l} \right )\textup{Pr} \left ( x_{l},\check{x}_{l} \right )
\end{equation} 
where $q$ is number of transmitted bits, $\textup{P}\left ( x_{l} \right )$ denotes the probability of $x_{l}$ and $e\left ( x_{l}\rightarrow  \check{x}_{l} \right )$ is the number of bit errors between $x_{l}$ and $\check{x}_{l}$.
\vspace{-0.3cm} 
\subsection{Asymptotic diversity order}

The achievable diversity order of NOMA users is obtained from the slope of the PEP at high SNR values, which can be evaluated numerically, using \eqref{eq:pep1}, \eqref{eq:PEP_L} or \eqref{eq:PEP_L1}, as follows
\begin{equation}
d_{s}= \lim_{\bar{\gamma}\rightarrow \infty } -\frac{\textup{log}\;\textup{Pr}\left ( x_{l}\rightarrow \check{x}_{l} \right )}{\textup{log}\;\bar{\gamma}}.
\end{equation}

\begin{figure}[ht] 
\centering
\includegraphics[width=0.7\linewidth]{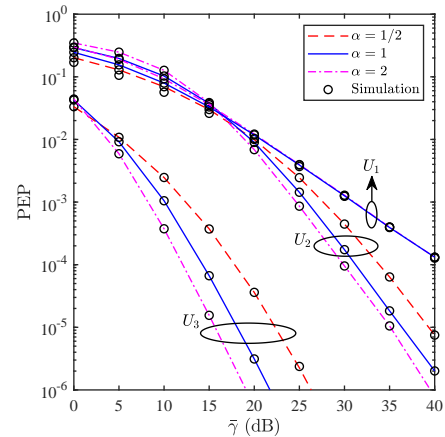}\vspace{-0.2cm}
\caption{The exact PEP of three users NOMA under different noise scenarios, i.e., $\alpha=1/2$, $\alpha=1$ and $\alpha=2$.}
\label{fig:alpha}
\end{figure}
\vspace{-0.8cm}
\begin{figure}[ht] 
\centering
\includegraphics[width=0.7\linewidth]{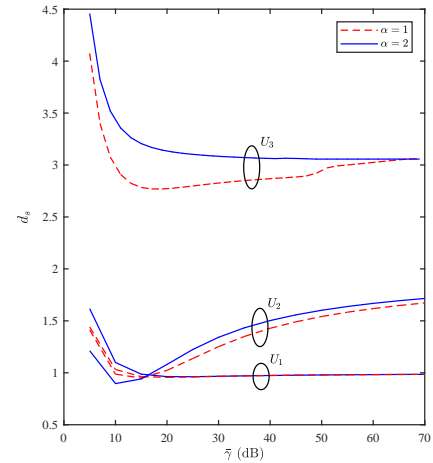}\vspace{-0.2cm}
\caption{The achievable diversity order of a three users NOMA under different noise scenarios, $\alpha=1$, $\alpha=2$.}
\label{fig:Div}
\end{figure}
\vspace{-0.2cm}
\begin{figure}[ht] 
\centering
\includegraphics[width=0.7\linewidth]{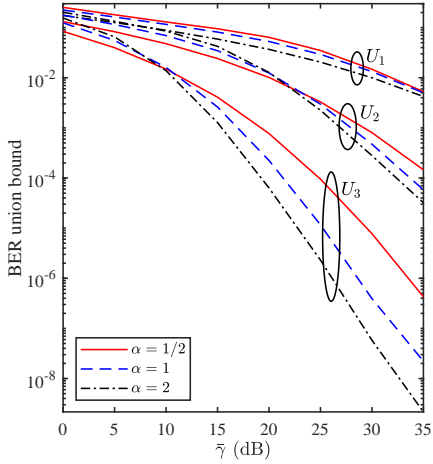}\vspace{-0.2cm}
\caption{BER union bound of a three users NOMA under different noise scenarios, $\alpha=1/2$, 1 and 2.\vspace{-0.2cm}}
\label{fig:UB}
\end{figure}
\vspace{-0.2cm}
\section{Numerical Results} \vspace{-0.1cm}
\label{sec:result}

In this section, we assess the accuracy of our analytical results presented in Proposition \ref{theorem:pep} when compared with Monte Carlo simulations. Without loss of generality, we consider a downlink NOMA system with a single BS and three users, i.e., $U_{1}, U_{2}$ and $U_{3}$. The transmitted and detected signals are selected randomly from a BPSK constellation. Power allocation coefficients are selected as the following, $a_{1}=0.7$, $a_{2}=0.2$ and $a_{3}=0.1$. 

Fig. \ref{fig:alpha} shows the analytical and simulated PEP of all users versus the average transmit SNR, $\bar{\gamma}$. The perfect match between the analytical and simulation results show the accuracy of the derived expression in \eqref{eq:pep1}. Moreover, it can be noticed that $\alpha$ has no effect on the PEP performance of the first user, given that the diversity order of the first user is limited to unity. Additionally, it is worth recalling that the first user does not perform SIC, consequently, it suffers from high interference from the second and third users. For the second and third users, it is observed that for small $\alpha$ values, the diversity order of the users is highly affected by $\alpha$, and as the value of $\alpha$ increases, its effect on the diversity order becomes negligible. This is verified by Fig. \ref{fig:Div} which presents the achievable diversity order of all users for $\alpha=1$ and 2. From Fig. \ref{fig:Div} it is shown that the diversity order of the $l$th user converges to $l$, for all users. 

The BER union bound is presented in Fig. \ref{fig:UB}, where the PEP is averaged over all possible scenarios of transmitted and detected symbols of all users \cite{8501953}. Fig. \ref{fig:UB} further corroborates the effect of $\alpha$ on the BER performance and the diversity order of NOMA users. It can be noted from the figure that higher order users are more susceptible to the variation of $\alpha$ values. More specifically, it is shown that $\alpha$ has negligible effect on the BER performance of the first user. This is due to the fact that the BER performance of the first user is dominated by the interference caused by all other users. Meanwhile, for higher order users, it can be observed that as $\alpha$ increases, its effect on the error rate performance decreases.   

\vspace{-0.3cm}
\section{Conclusions}
\label{sec:conc}

In this letter, we investigated the  performance of NOMA systems subject to AWGGN with imperfect SIC. Particularly, we derived an exact expression for the PEP in the considered setup, and studied the impact of the GGN parameters on the PEP performance and the achievable diversity order. From both simulation and analytical results, for small values of $\alpha$, i.e., $\alpha<1$, a noticeable impact on the diversity order of NOMA users was observed. On the other hand, for larger values of $\alpha$, it was noticed that the achievable diversity order is independent of $\alpha$. Hence, as in the AWGN case, the diversity order is dominated by the user's order.

\vspace{-0.3cm}
\bibliographystyle{IEEEtran}
\bibliography{references}

\end{document}